\documentclass[preprint,12pt,showpacs,aps,amsmath,amssymb,
nofootinbib,superscriptaddress,showkeys]{revtex4}

\usepackage{epsfig}
\usepackage{subfigure}
\usepackage{amsmath}
\usepackage{amssymb}
\usepackage{bbm}
\usepackage{wasysym}
\usepackage{mathrsfs}
\usepackage{graphicx}
\newcommand{\beq}{\begin{equation}}
\newcommand{\eeq}{\end{equation}}
\newcommand{\beqa}{\begin{eqnarray}}
\newcommand{\eeqa}{\end{eqnarray}}

\newcommand{\nn}{\nonumber}
\newcommand{\dmsq}{\Delta m^2}
 \newcommand{\dma}{{\Delta m}^2_{32}}
 \newcommand{\dms}{{\Delta m}^2_{21}}
\begin{document}

\title{Renormalization group evolution of neutrino mixing parameters
near $\theta_{13} = 0$ and models with vanishing $\theta_{13}$ 
at the high scale}

\author{Amol Dighe}
\email{amol@theory.tifr.res.in}
\affiliation{Tata Institute of Fundamental Research, 
Homi Bhabha Road, Colaba, Mumbai 400005, India}

\author{Srubabati Goswami}
\email{sruba@prl.res.in}
\affiliation{Physical Research Laboratory, Navrangpura,
Ahmedabad 380009, India}

\author{Shamayita Ray}
\email{shamayitar@theory.tifr.res.in}
\affiliation{Tata Institute of Fundamental Research,
Homi Bhabha Road, Colaba, Mumbai 400005, India}


\begin{abstract}
Renormalization group (RG) evolution of the neutrino mass matrix may
take the value of the mixing angle $\theta_{13}$ very close to
zero, or make it vanish. On the other hand, starting from $\theta_{13}=0$
at the high scale it may be possible to generate a non-zero $\theta_{13}$
radiatively.  
In the most general scenario with non-vanishing CP violating
Dirac and Majorana phases, we explore the evolution in the 
vicinity of $\theta_{13}=0$, in terms of its structure in the
complex ${\cal U}_{e3}$ plane.
This allows us to explain the apparent singularity in the
evolution of the Dirac CP phase $\delta$ at $\theta_{13}=0$. 
We also introduce a formalism for calculating the RG evolution
of neutrino parameters that uses the Jarlskog invariant and
naturally avoids this singular behaviour.
We find that the parameters need to be extremely fine-tuned 
in order to get exactly vanishing $\theta_{13}$ during evolution.
For the class of neutrino mass models with $\theta_{13}=0$ at the
high scale, we calculate the extent to which RG evolution can
generate a nonzero $\theta_{13}$, 
when the low energy effective theory is the standard 
model or its minimal supersymmetric extension.
We find correlated constraints on $\theta_{13}$, the lightest 
neutrino mass $m_0$, the effective
Majorana mass $m_{ee}$ measured in the neutrinoless double 
beta decay, and the supersymmetric parameter $\tan\beta$.
\end{abstract}



\preprint{TIFR/TH/08-43}

\maketitle


\section{Introduction}
\label{sec-intro}

In the last decade, neutrino experiments have reached a stage where
the basic structure of the neutrino masses and mixing is more or less
clear.  
We know that the three neutrino flavors ($\nu_\alpha, \alpha \in
\{e,\mu,\tau\}$) mix to form three neutrino mass eigenstates
($\nu_i, i \in \{1,2,3\}$),
which are separated by
$\dmsq_{ij} \equiv m_i^2 - m_j^2$ where, $m_{i,j}$ denote mass 
eigenvalues  with $ i,j \in \{ 1,2,3 \}$.
The two sets of eigenstates are connected through 
$\nu_\alpha = (U_{\rm PMNS})_{\alpha i} \nu_i$,
where $U_{\rm PMNS}$ is the Pontecorvo-Maki-Nakagawa-Sakata 
neutrino mixing matrix \cite{pontecorvo,mns} in the basis where 
the charged lepton mass matrix is assumed to be diagonal. 
This matrix is parametrized as
\beq
U_{\rm PMNS} =  
\left( \begin{array}{ccc}
e^{i \chi_1} & 0 & 0 \\
0 & e^{i \chi_2} & 0 \\
0 & 0 & e^{i \chi_3} \\
\end{array} \right)
\cdot {\cal U}  \cdot
\left( \begin{array}{ccc}
e^{i \phi_1} & 0 & 0 \\
0 & e^{i \phi_2} & 0 \\
0 & 0 & 1 \\
\end{array} \right) \; ,
\label{Upmns}
\eeq
where ${\cal U}$ is the matrix 
\beq
{\cal U} =
\left( \begin{array}{ccc}
c_{12}c_{13} & s_{12}c_{13} & s_{13}e^{-i\delta} \\
-s_{12}c_{23}-c_{12}s_{23}s_{13}e^{i\delta}  
& c_{12}c_{23}-s_{12}s_{23}s_{13}e^{i\delta} & s_{23}c_{13} \\
s_{12}s_{23}-c_{12}c_{23}s_{13}e^{i\delta} & 
-c_{12}s_{23}-s_{12}c_{23}s_{13}e^{i\delta} & c_{23}c_{13} \\
\end{array} \right) \; .
\label{Uckm}
\eeq
Here $c_{ij}$ and $s_{ij}$ are the cosines and sines respectively
of the mixing angle $\theta_{ij}$, 
$\delta$ is the 
Dirac CP violating phase, $\phi_i$ are the Majorana phases, 
and $\chi_i$ are the so-called unphysical phases that do not
play a role in the phenomenology of neutrino mixing, 
but whose values may be predictable within the context 
of specific models.
The current best-fit values and 3$\sigma$ ranges for these 
parameters are summarized in Table~\ref{tab:bounds}. 
It is not known whether the neutrino mass ordering 
is normal ($m_1 < m_2 < m_3$) 
or inverted ($m_3 < m_1 < m_2$).
\begin{table}[h]
\begin{center}
\begin{tabular}{ccccc}\hline
& \phantom{space} & Best fit & \phantom{space} & $3\sigma$ range   \\\hline
$\Delta m_{21}^2$ [$10^{-5}{\rm eV}^2$] & & 7.65 & & 7.05 - 8.34 \\
$|\Delta m_{31}^2|$ [$10^{-3}{\rm eV}^2$] & & 2.40 && 2.07 - 2.75 \\\hline
$\sin^2\theta_{12}$ && 0.304 && 0.25 - 0.37 \\
$\sin^2\theta_{23}$ && 0.50 && 0.36 - 0.67 \\
$\sin^2\theta_{13}$ && 0.01 && $\leq$ 0.056 \\\hline
\end{tabular}
\caption{The present best-fit values and 3$\sigma$ ranges
of oscillation parameters \cite{Schwetz:2008er}.
\label{tab:bounds}}
\end{center}
\end{table}

An intriguing situation with the neutrino mixing is that
two of the mixing angles, $\theta_{12}$ and $\theta_{23}$, are
definitely large, while the third angle $\theta_{13}$ is small
and may even be zero.
Such a situation is indicative of some kind of symmetry principle
at work. 
Indeed, there is a whole class of models 
with $\theta_{13} \approx 0$ that are consistent with 
data \cite{albright}. 
$\theta_{23} = \pi/4$ and $\theta_{13} = 0$ are 
allowed by the current data and their origin  
has been traced to an exact $\mu-\tau$ exchange symmetry in 
the neutrino mass matrix \cite{mutau}. 
Such symmetries can be realized 
by models based on the discrete non-abelian
symmetry groups like 
$A_4$ \cite{a4},  $D_4$ \cite{d4}, $S_3$ \cite{s3}, $S_4$ \cite{s4}. 
Special cases of an exact $\mu-\tau$ symmetric matrix 
corresponding to a $L_e$ symmetry for normal ordering \cite{le}, 
$L_e -L_\mu -L_\tau$ symmetry for inverted ordering \cite{le-lmu-ltau} 
and $L_\mu - L_\tau$ 
symmetry for quasidegenerate neutrinos can give
$\theta_{13}=0$ \cite{lmu-ltau}. Any deviation from  this value 
would indicate breaking of these symmetries. 
Models with discrete abelian symmetries 
can also make $\theta_{13}$ vanish
\cite{low}. Models involving
certain texture zeroes in the neutrino 
Yukawa matrix  or certain scaling relations 
between Majorana matrix elements 
can also predict zero or  almost vanishing 
$\theta_{13}$  \cite{texture}.
SO(10) models with certain structures for Dirac mass 
matrices \cite{so10}, or those with a SO(3) symmetry can predict
$\theta_{13} \lesssim 10^{-4}$ with a normal mass ordering \cite{so3}.

Most of the symmetries in these models are obeyed at the high scale, 
and are broken at the low scale by, for example, radiative corrections.
If the radiative corrections are large enough, any trace of the
original symmetry may be wiped out. 
However in the context of a specific model, the compatibility
between the high scale symmetry and low scale measurements can
still be verified.
This needs a careful study of the renormalization group (RG) 
evolution of the neutrino mass matrix and the mixing parameters.
The basic formalism for calculating this evolution
has been established in 
\cite{babu-pantaleone,chankowski-pokorski,antusch,antusch-HDM-MSSM}.
Specific features of the evolution, like the stability of 
mixing angles and masses 
\cite{ellis-lola,haba-stability,ma-stability,Haba:2000rf}, 
possible occurrence of fixed points 
\cite{chankowski-fixed,Pantaleone-fixed,xing-fixed},
evolution of nearly degenerate Majorana neutrinos 
\cite{vissani-deg,branco-deg,haba-deg,casas-deg,adhikari-deg,
Joshipura:2002xa,Joshipura:2003fy,xing-deg,petcov-majorana}, 
the generation of large mixing angles
from small angles at the high scale 
\cite{tanimoto,haba-LA,balaji-dighe, Balaji:2000ma,mpr,Agarwalla:2006dj},
or radiative generation of ${\cal U}_{e3}$ starting 
from zero value at high scale 
\cite{Joshipura:2002kj,Joshipura:2002gr,Mei:2004rn}, 
have been explored.
Threshold effects on masses and mixings, due to the decoupling
of heavy particles involved in the neutrino mass generation, 
have also been estimated 
\cite{chankowski-threshold,antusch-threshold,Mohapatra:2005gs}.
These effects can revive \cite{antusch-LMA,shindou-LMA}
the bimaximal mixing scenario \cite{bimaximal}, which 
predicts $\theta_{13}=0$. 

Analytical expressions for the RG evolution of these parameters
have been obtained through an expansion in the small parameter 
$\theta_{13}$ \cite{antusch}.
For a quantity $X \in \{m_i, \theta_{ij}, \phi_i \}$, 
the evolution may be written as
\beq
\dot{X} = A_X + 
 {\cal O}(\theta_{13}) \; ,
\label{dotX}
\eeq
where dot represents the derivative with respect to 
$t \equiv \ln(\mu/{\rm GeV}) / (16 \pi^2)$, with
$\mu$  the relevant energy scale.
Here $A_X$ is independent of $\theta_{13}$, but is a function  of
$m_i,\theta_{12},\theta_{23},\phi_i,\delta$ in general.
In the context of quark-lepton complementarity, approximate but
transparent analytical expressions were obtained
in \cite{rgqlc} where a further expansion in the small parameter
$\Delta_\tau \propto y_\tau^2 (1 + \tan^2 \beta)$ was employed. 
Here $y_\tau$ is the Yukawa coupling of the tau lepton 
and $\tan\beta$ the ratio of vacuum expectation values of the two Higgses
in minimal supersymmetric standard model (MSSM).   
Such an expansion was used to constrain the allowed values of
mixing angles in the context of tri-bimaximal mixing \cite{tbm-planck}
and to distinguish between various symmetry-based relations 
at the high scale by comparing the low scale $\theta_{13}$
values \cite{qlctbm}.

A subtle but important issue arises in the evolution of the Dirac
phase $\delta$ at $\theta_{13}=0$. 
With the parametrization in \cite{antusch},
the evolution formally takes the form 
\beqa
\dot{\delta} &=& \frac{D_\delta}{\theta_{13}} + A_\delta +
{\cal O}(\theta_{13}) \; , \label{delta-dot-intro}
\eeqa
such that the derivative of $\delta$ formally diverges at vanishing
$\theta_{13}$, indicating an apparent singularity.
This is an unphysical singularity: all the elements of the mixing matrix 
$U_{\rm PMNS}$ evolve continuously, and the peculiar evolution of $\delta$
is related to the fact that $\delta$ is undefined at $\theta_{13}=0$.
This argument is in fact
used in \cite{antusch} to assert that $D_\delta$ identically vanishes 
when $\theta_{13}=0$, which leads to a specific value of $\cot\delta$
which is a function of $\{m_i,\phi_i\}$ at  $\theta_{13}=0$. 
Ref.~\cite{xing-fixed} has examined this prescription in various 
limits in the parameter space.

While the above prescription for choosing the value of $\delta$ 
at $\theta_{13}=0$ works practically when one needs to start with 
vanishing $\theta_{13}$, a few conceptual problems remain.
Firstly, when $\theta_{13}=0$, the value of $\delta$ chosen should
not make a difference to the RG evolution since $\delta$ is an
unphysical quantity at this point.
Secondly, it is not {\it a priori} clear whether the prescription
would work when $\theta_{13} =0$ is reached during the process of 
RG evolution.
Indeed, getting the required value of $\delta$ precisely when 
$\theta_{13}=0$ 
may seem like fine tuning.
The prescription in \cite{antusch}, though practical, does not tell us 
the origin of this apparent coincidence.
Here we analyze this problem in more detail, and find an explanation
in terms of the evolution of the complex quantity 
${\cal U}_{e3} = \sin \theta_{13} e^{-i\delta}$ in the parameter plane
Re$({\cal U}_{e3})$--Im$({\cal U}_{e3})$.

We also evolve an alternative formalism
where the singularity does not arise at all. This is based on the 
observation that the set of quantities 
${\cal P}_J \equiv 
\{m_i, \theta_{12}, \theta_{23}, \theta_{13}^2, \phi_i, J_{\rm CP}, J'_{\rm CP}\}$,
where $J_{\rm CP} = \frac{1}{2} s_{12} c_{12} s_{23} c_{23} s_{13} c_{13}^2 \sin\delta$ 
is the Jarlskog invariant and 
$J'_{\rm CP} = \frac{1}{2} s_{12} c_{12} s_{23} c_{23} s_{13} c_{13}^2 \cos\delta$, 
have the same information
as the set ${\cal P}_\delta \equiv 
\{m_i, \theta_{12}, \theta_{23}, \theta_{13}, \phi_i, \delta \}$. 
We therefore write the evolution equations in terms of the former 
set and explicitly show that the complete evolution may be 
studied without any reference to diverging quantities.
We confirm numerically that the evolutions with both the parametrizations
indeed match with each other and with the exact numerical one.

With the conceptual issue clarified, we numerically study the
extent to which $\theta_{13}$ may be generated through RG running
in the class of models with $\theta_{13}=0$ at the high scale,
where the low energy effective theory is the standard model (SM) or
the minimal supersymmetric standard model (MSSM).
This evolution turns out to be extremely sensitive to the mass
of the lightest neutrino $m_0$, the neutrino mass ordering and
the Majorana phases. 
Another experimentally observable quantity that
depends on these parameters is the effective Majorana mass
$m_{ee}$ which is explored by the neutrinoless double beta decay 
experiments. Correlated constraints can therefore be obtained on
$\theta_{13}$, $m_0$ and $m_{ee}$, the quantities for which only upper
bounds are available currently but which may be measured in the
next generation experiments. For the case of MSSM, it will also 
depend on the value of $\tan\beta$.

The paper is organized as follows. Sec.~\ref{singularity} deals
with the apparent singularity in the evolution of $\delta$.
Sec.~\ref{new-param} calculates the RG evolution in terms
of the parameter set ${\cal P}_J$.
Sec.~\ref{bounds} determines the upper bounds on the value of
$\theta_{13}$ generated through the RG evolution in the SM
and the MSSM. In Sec.~\ref{summary}, we summarize our results.

\section{Apparent singularity in $\dot{\delta}$ at $\theta_{13}=0$ and 
RG evolution in the complex ${\cal U}_{e3}$ plane}
\label{singularity}

Analytic studies of the evolution of neutrino parameters
till date have been mostly performed with the parameter set
${\cal P}_\delta \equiv 
\{m_i, \theta_{12}, \theta_{23}, \theta_{13}, \phi_i, \delta \}$. 
The RG evolution equations obtained are all continuous and non-singular, 
except the equation for the Dirac CP phase $\delta$, which is given by 
\beqa
\dot{\delta} &=&  \frac{D_\delta}{\theta_{13}} + A_{\delta} + 
{\cal O}(\theta_{13}) \; , \label{delta-dot}
\eeqa
where
\beqa
D_\delta &=& \frac{C y_\tau^2}{2} \sin{2 \theta_{12}} \sin{2 \theta_{23}} 
\frac{m_3 }{\dmsq_{31}} \times \nn \\
&& \Bigl[  m_1 \sin{(2 \phi_1 - \delta)} 
- (1 + \zeta) m_2 \sin{(2 \phi_2 - \delta)} + \zeta m_3 \sin{\delta} \Bigr] \; ,
\label{D-delta}
\eeqa 
\beqa
A_{\delta} &=& 2 C y_\tau^2 \Biggl[ 
\frac{m_1 m_2 }{\dms} s_{23}^2 \sin{(2 \phi_1 - 2 \phi_2)} \nn \\ 
&& + \frac{m_1 m_3}{\dmsq_{31}} \left( c_{12}^2 c_{23}^2 
\sin{(2 \delta - 2\phi_1)} + s_{12}^2 \cos{2 \theta_{23}} \sin{2\phi_1} \right) \nn \\
&& + \frac{m_2 m_3}{\dma} \left( s_{12}^2 c_{23}^2 
\sin{(2 \delta - 2\phi_2)} + c_{12}^2 \cos{2 \theta_{23}} \sin{2\phi_2} \right)\Biggr] \; .
\label{A-delta}
\eeqa
Here $\zeta = \dms/ \dma$ and $C$ is a constant which 
depends on the underlying effective theory in the energy regime considered. 
Eq.~(\ref{delta-dot})
clearly suggests that $\dot{\delta}$ diverges for $\theta_{13} \to 0$. 
This problem is overcome by requiring that $D_\delta = 0$ at 
$\theta_{13} = 0$, which gives the 
following condition on $\delta$ at $\theta_{13} = 0$ \cite{antusch}:
\beqa
\cot{\delta} = 
\frac{m_1 \cos{2\phi_1} - (1+\zeta) m_2 \cos{2 \phi_2} - \zeta m_3}
{m_1 \sin{2\phi_1} - (1+\zeta) m_2 \sin{2 \phi_2}} \; .
\label{cot-delta}
\eeqa
The above prescription works for the calculation of evolution
when one starts with vanishing $\theta_{13}$. However
on the face of it, it seems to imply that 
the CP phase $\delta$, which does not have any physical meaning 
at the point $\theta_{13} = 0$, should attain a particular value depending 
on the masses and Majorana phases, as given in eq.~(\ref{cot-delta}). 
Also, the situation when $\theta_{13}=0$ is reached during the
course of the RG evolution has not been studied so far,
so it is not clear if the prescription needs to be introduced
by hand in such a case, or whether the RG evolution equations
stay valid while passing through $\theta_{13}=0$.
Getting the required value of $\delta$ precisely when 
$\theta_{13}=0$ would seem to need fine tuning, unless we are able to figure
out the origin of this apparent coincidence, and  show
that this value of $\delta$ is a natural limit of the RG evolution.

The problem also propagates to the evolution of $\theta_{13}$,
since it depends in turn on $\delta$:
\beqa
\dot{\theta}_{13} &=& A_{13} +
{\cal O}(\theta_{13}) 
\label{th13-dot} \; ,\\
A_{13} &=& \frac{C y_\tau^2}{2} \sin{2 \theta_{12}} \sin{2 \theta_{23}} 
\frac{m_3}{\dmsq_{31}} \times \nn \\
&& \left[ m_1 \cos{(2 \phi_1 - \delta)} - (1 + \zeta) m_2 \cos{(2 \phi_2 - \delta)} 
- \zeta m_3 \cos{\delta} \right] \label{A13} \; .
\eeqa
The evolution of all the other quantities, 
viz. $\theta_{12}, \theta_{23}, m_i, \phi_i$ is 
independent of $\delta$ upto ${\cal O}(\theta_{13}^0)$
\cite{antusch}, so these quantities do not concern us here.

\begin{figure}
\parbox{3in}{
\epsfig{file=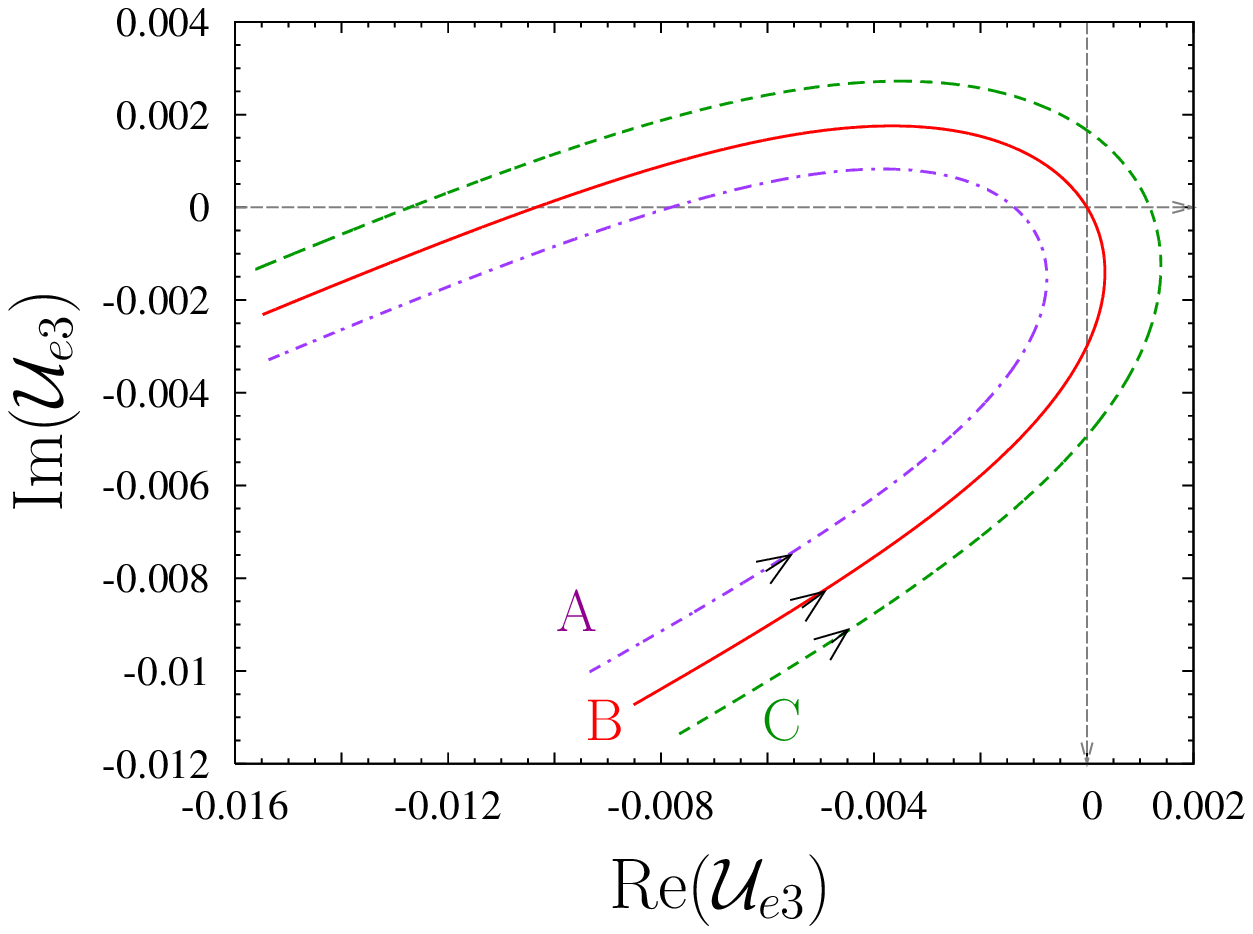,width=3in}
}
\parbox{3in}{
\epsfig{file=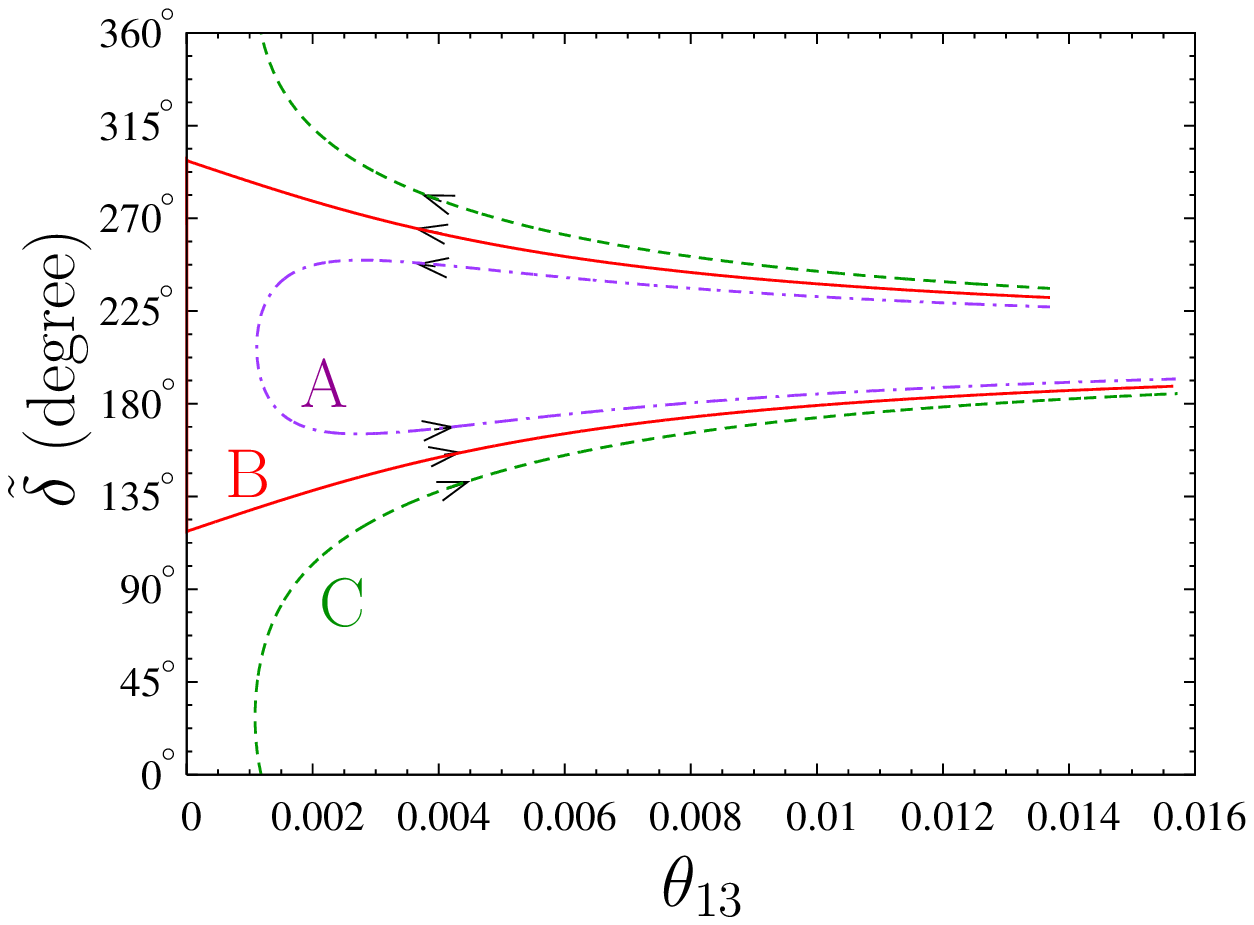,width=3in}
}
\caption{The left panel shows the evolution 
in the Re$({\cal U}_{e3})$--Im$({\cal U}_{e3})$ parameter plane, whereas 
The right panel shows the corresponding evolution in
the $\theta_{13}$--$\tilde\delta$ plane.
The values of the parameters chosen at $\mu_0 = 10^{12}$ GeV are:
$\tan{\beta} = 50$, $m_0 = 0.0585$ eV$^2$, 
$\dms = 4.22 \times 10^{-5}$ eV$^2$, $\dma = 3.91 \times 10^{-3}$ eV$^2$, 
$\theta_{12} = 32.84^\circ$, $\theta_{23} = 43.71^\circ$ and 
$\theta_{13} = 0.014$ rad.
The Majorana phases are taken to be $\phi_1 = 58.9^\circ$ 
and $\phi_2 = 159.15^\circ$. 
The Dirac CP phase is $124.0 ^\circ$ for case A (violet, dash-dotted line), 
$128.447^\circ$ for case B (red, solid line) and 
$133.0^\circ$ for case C (green, dashed line).
\label{fig:argand}
}
\end{figure}

In order to understand the nature of the apparent singularity
in $\delta$, we explore the
RG evolution of the complex quantity 
${\cal U}_{e3} = \sin \theta_{13} e^{-i\delta}$.
We start with three representative values of $\delta$ at the
energy scale $\mu_0 = 10^{12}$ GeV, with the other parameters chosen
such that $\theta_{13} \lesssim 10^{-3}$ at $\mu \approx 10^{9}$ GeV.
The left panel of Fig.~\ref{fig:argand} shows the evolution 
in the complex ${\cal U}_{e3}$ plane.
The right panel shows the corresponding evolution in
the $\theta_{13}$--$\tilde \delta$ plane, 
with $\tilde \delta \equiv 2\pi - \delta$. 
The following observations may be
made from the figures:\\
(a) Though all the parameter values at the high scale are very close,
and though in all cases $\theta_{13}$ decreases to a very small
value before it starts to increase, $\theta_{13}$ does not vanish
during the evolution in all the cases.
Indeed, the value of $\delta$  chosen at the high scale,
in order to make $\theta_{13}$ vanish during its evolution, needs
to be extremely fine-tuned. This is because 
\beq
\sin^2 \theta_{13} = [ {\rm Re}({\cal U}_{e3})]^2 + 
[{\rm Im}({\cal U}_{e3})]^2 \; ,
\eeq
so that one needs both the real and imaginary components of ${\cal U}_{e3}$
to vanish simultaneously, which needs a coincidence.
Note that when both the CP violating phases $\delta$ and $\phi_i$ vanish
at the high scale,
Im$({\cal U}_{e3}) =0$ automatically throughout the evolution. 
Then starting from a non-zero value at high scale, $\theta_{13}$ can be 
made to vanish simply by requiring Re$({\cal U}_{e3}) =0$ so that 
no fine tuning is needed. \\
(b) 
With the definition $\tilde\delta \equiv  2\pi -\delta$ 
we have ${\cal U}_{e3} \equiv s_{13} e^{i \tilde{\delta}}$ and 
thus $\tilde\delta$ is the phase of ${\cal U}_{e3}$ which can
be read off easily from the Re$({\cal U}_{e3})$--Im$({\cal U}_{e3})$ plot.
The values of $\delta$ chosen at $\mu_0 = 10^{12}$ GeV are such
that $\tilde \delta$ is in the third quadrant, 
so Re$({\cal U}_{e3}) <0$ and Im$({\cal U}_{e3})<0$ at this scale.  
At the end of the evolution, at $\mu=10^{4}$ GeV, 
$\tilde\delta$ returns to the third quadrant.
During its evolution, $\tilde\delta$ may change its quadrant
zero, one or multiple times.
The value of $\theta_{13}$ need not vanish completely during 
the RG evolution, as is represented by the scenarios A and C.
Scenario B is the one where Re$({\cal U}_{e3})$ and Im$({\cal U}_{e3})$
vanish at the same point, and therefore $\theta_{13}$ passes through 
zero during its evolution. \\
(c)
In scenario A, since Re$({\cal U}_{e3})$ stays negative, 
$\tilde\delta$ simply moves from the third quadrant 
to the second, and then returns to the third in a continuous manner.
In scenario C on the other hand, $\tilde\delta$ has to 
pass through the fourth, first and second quadrant in 
sequence to finally return to the third quadrant. 
However its evolution is continuous, the 
apparent jump at the lowest $\theta_{13}$ values 
in the right panel of Fig.~\ref{fig:argand} is just the 
identification of $0^\circ$ and $360^\circ$. \\
(d)
In scenario B, $\tilde\delta$ starts in the third quadrant and moves 
continuously 
to the fourth quadrant. However it propagates to the second quadrant
directly through the origin, thus bypassing the first
quadrant entirely. Its value at the origin can be well-defined through
the limit
\beq
\cot \tilde\delta_0 \equiv 
\lim_{ {\rm Re}({\cal U}_{e3}), {\rm Im}({\cal U}_{e3}) \to 0} 
\quad
\frac{ {\rm Re}({\cal U}_{e3})}{{\rm Im}({\cal U}_{e3})} =
\lim_{ {\rm Re}({\cal U}_{e3}), {\rm Im}({\cal U}_{e3}) \to 0} 
\quad
\frac{ \frac{d}{dt} {\rm Re}({\cal U}_{e3})}{ \frac{d}{dt}
{\rm Im}({\cal U}_{e3})}
\eeq
where we have used L'Hospital's rule to compute the limit since
both the numerator and denominator in this ratio tend to zero at the 
limiting point.

Since
\beq
{\rm Re}({\cal U}_{e3}) = \sin \theta_{13} \cos {\delta} \; , \quad
{\rm Im}({\cal U}_{e3}) = -\sin \theta_{13} \sin {\delta} \; , 
\eeq
we have
\beq
\cot \tilde\delta_0 =- \frac{A_{13} \cos\delta - 
D_\delta \sin\delta}{A_{13} \sin\delta+ D_\delta \cos\delta} \; ,
\eeq
and using eqs.~(\ref{D-delta}) and (\ref{A13}), one obtains
\beq
\cot \tilde\delta_0
= - \frac{m_1 \cos{2 \phi_1} - (1 + \zeta) m_2 \cos{ 2 \phi_2} - \zeta m_3}
{m_1 \sin{2 \phi_1} - (1 + \zeta) m_2 \sin{ 2 \phi_2}} \; .
 \eeq
Since $\delta= 2 \pi -\tilde\delta$, this is equivalent to
\beq
\cot \delta_0 = 
\frac{m_1 \cos{2\phi_1} - (1+\zeta) m_2 \cos{2 \phi_2} - \zeta m_3}
{m_1 \sin{2\phi_1} - (1+\zeta) m_2 \sin{2 \phi_2}} \; ,
\label{cot-delta0}
\eeq
which corresponds exactly to the value of $\cot\delta$ in 
eq.~(\ref{cot-delta}), which had been prescribed in \cite{antusch}.
We have thus shown that the prescription follows directly
from the procedure of taking the limit of $\delta$ as 
Re$({\cal U}_{e3})$ and Im$({\cal U}_{e3})$ go to zero simultaneously.

\begin{figure}
\parbox{3in}{
\epsfig{file=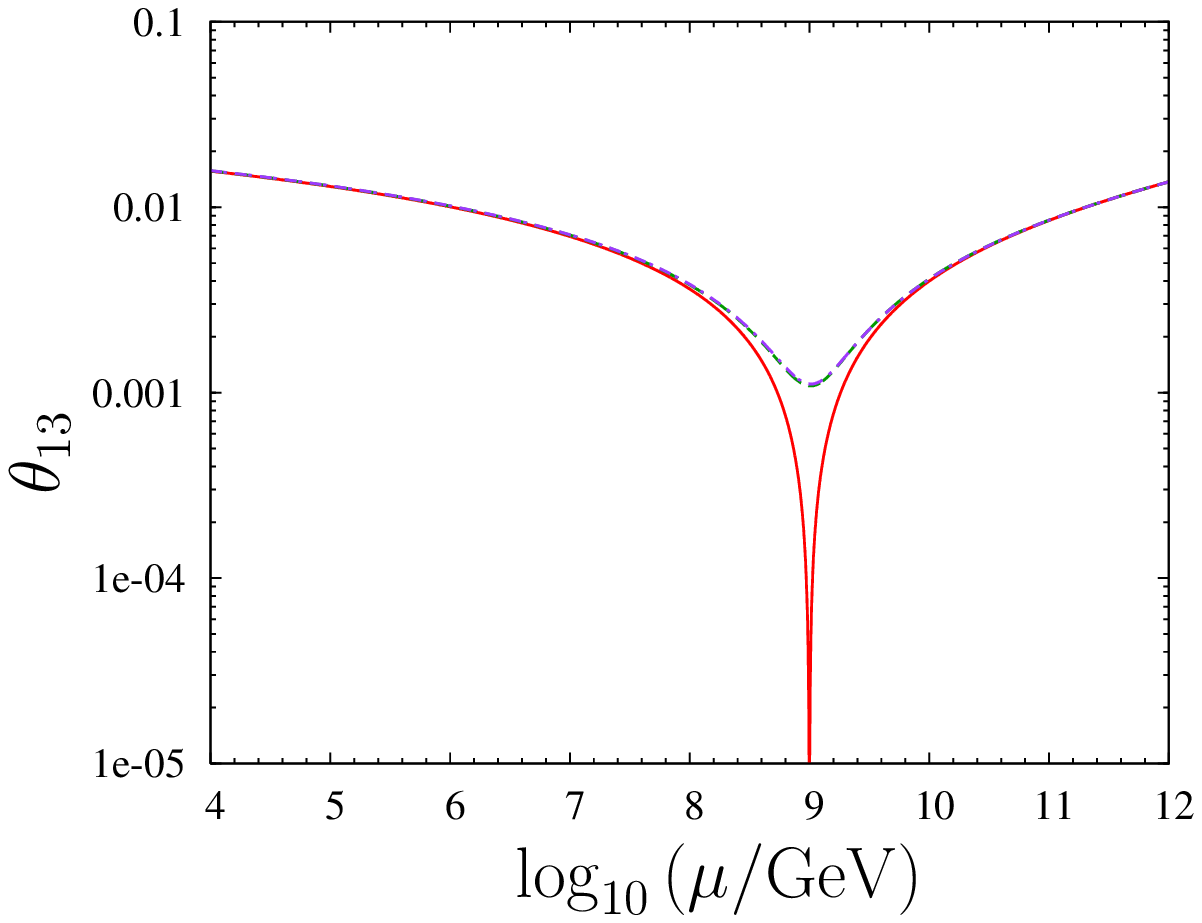,width=3in}
}
\parbox{3in}{
\epsfig{file=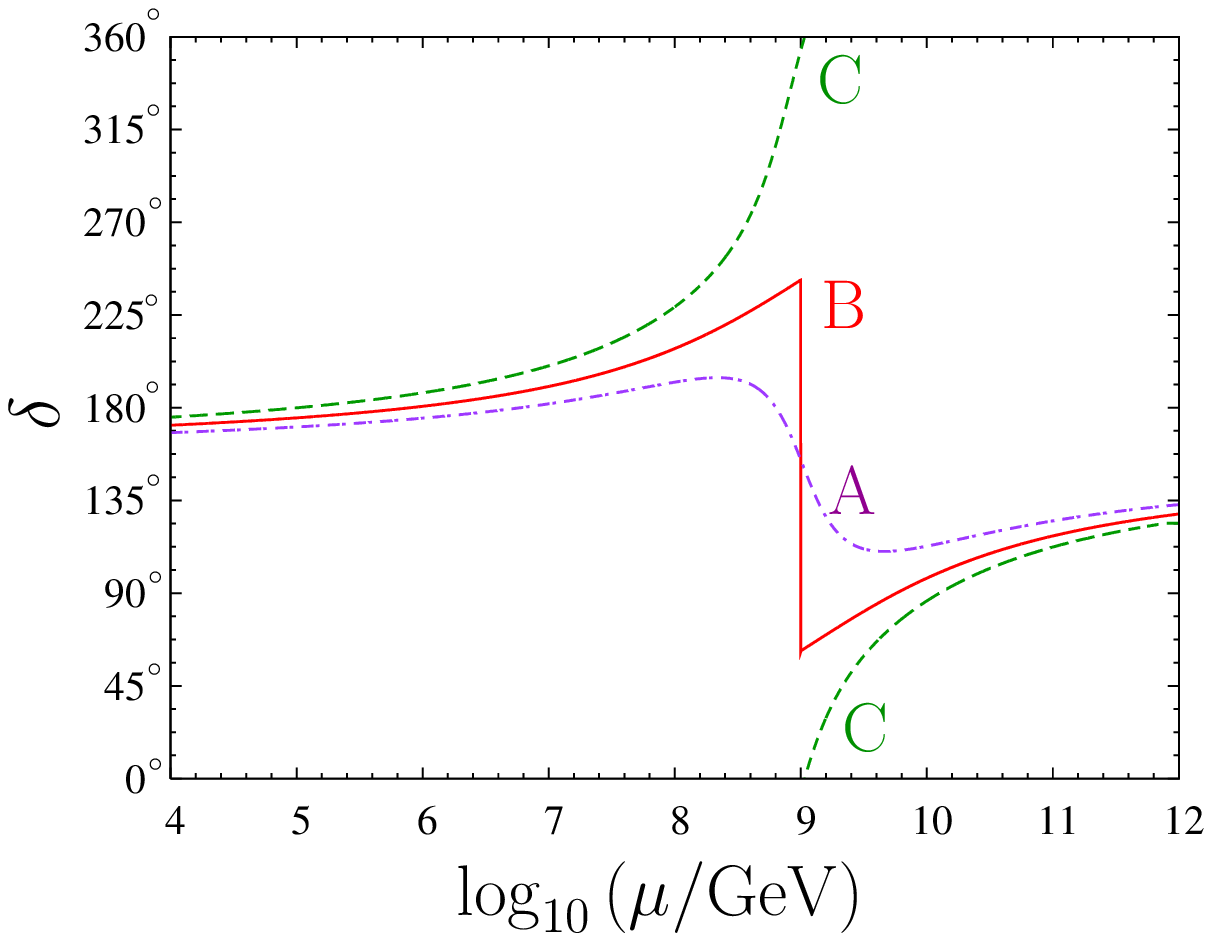,width=3in}
}
\parbox{3in}{
\epsfig{file=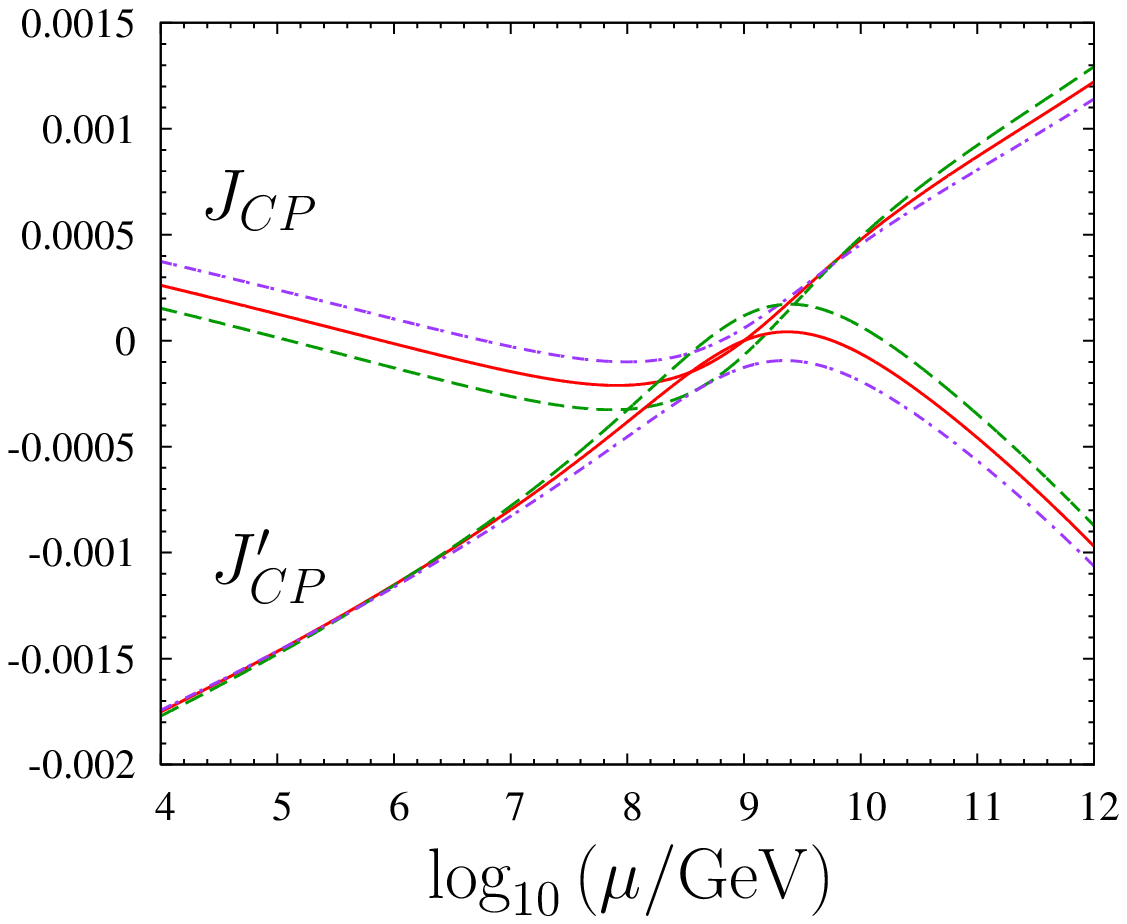,width=3in}
}
\parbox{3in}{
\epsfig{file=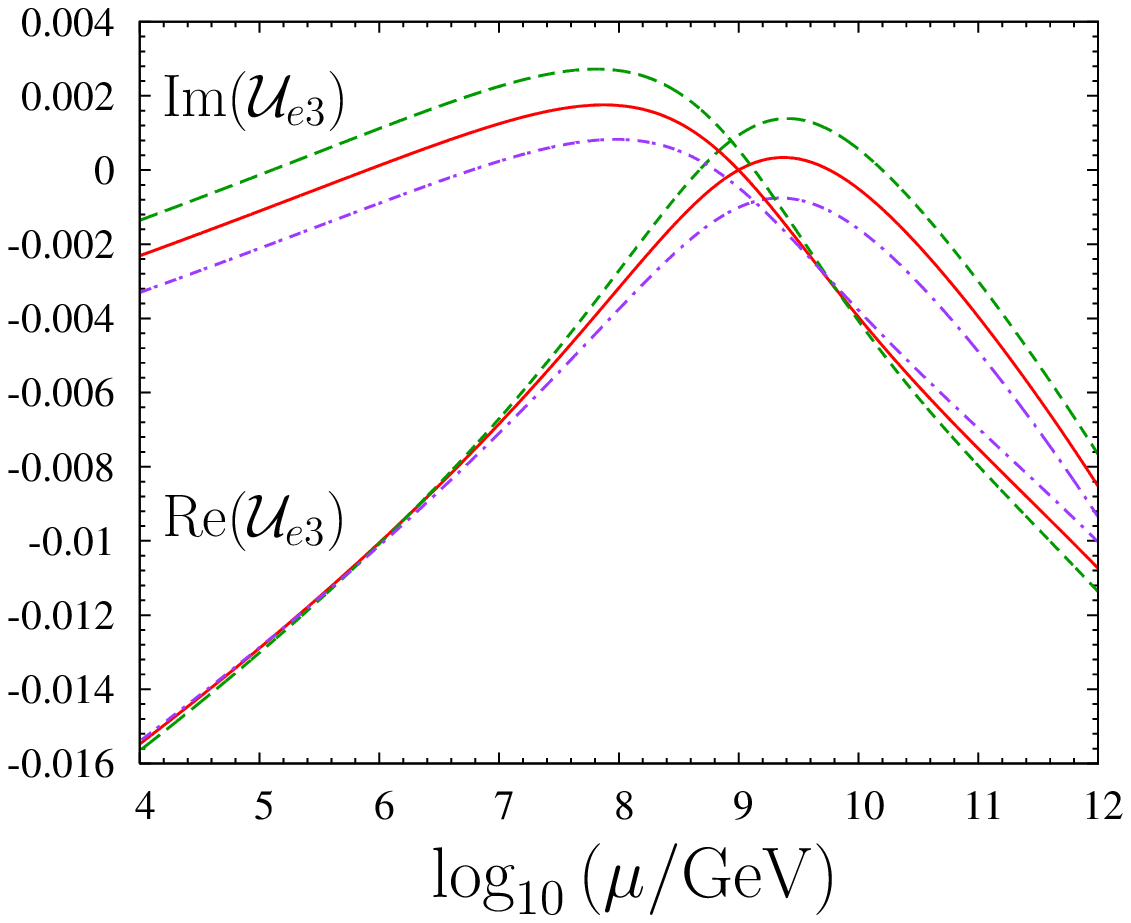,width=3in}
}
\caption{The evolution of $\theta_{13}, \delta$, Re$({\cal U}_{e3})$,
Im$({\cal U}_{e3})$, $J_{\rm CP}$ and $J'_{\rm CP}$ as functions of the energy scale 
$\mu$, in the scenarios 
A (violet, dash-dotted line), 
B (red, solid line) and 
C (green, dashed line).
\label{evol-fig}
}
\end{figure}

The net evolution of $\theta_{13}$ and $\delta$ as functions of
the energy scale has been shown in the top panels of
Fig.~\ref{evol-fig}.
The evolution of $\delta$ clearly has a discontinuity at $\theta_{13}=0$ 
in scenario B, where its value changes by $\pi$.
Though the origin of this discontinuity has now been well 
understood, it is important to have a clear evolution of parameters that
reflect the continuous nature of the evolution of elements
of the neutrino mixing matrix $U_{\rm PMNS}$.
This can clearly be achieved by using the parameters 
Re$({\cal U}_{e3})$ and Im$({\cal U}_{e3})$. However, we prefer to use the 
Jarlskog invariant 
\beq 
J_{\rm CP} \equiv \frac{1}{2} \sin{\theta_{12}}\cos{\theta_{12}} 
 \sin{\theta_{23}}\cos{\theta_{23}} \sin{\theta_{13}}\cos^2{\theta_{13}} 
\sin{\delta} \; ,
\label{jarlskog} 
\eeq
which appears in the probability expressions 
relevant for the neutrino oscillation experiments, and is
therefore more directly measurable than the real and imaginary 
parts of ${\cal U}_{e3}$. 
Since $J_{\rm CP}$ has information only about $\sin\delta$, we
need its partner
\beqa
J_{\rm CP}^\prime \equiv \frac{1}{2}  \sin{\theta_{12}}\cos{\theta_{12}} 
\sin{\theta_{23}}\cos{\theta_{23}} \sin{\theta_{13}}\cos^2{\theta_{13}} 
\cos{\delta} \; \label{J-prime} ,
\eeqa
to keep track of the quadrant in which $\delta$ lies.
The evolutions of $(J_{\rm CP},J'_{\rm CP})$ are very similar to those
of $({\rm Re}({\cal U}_{e3}), {\rm Im}({\cal U}_{e3}))$, as can be seen 
from the bottom panels of Fig.~\ref{evol-fig}.

\section{RG evolution equations in terms of the parameter
set ${\cal P}_J$}
\label{new-param}

We now calculate the RG evolution of the Jarlskog invariant
$J_{\rm CP}$ and its partner $J'_{\rm CP}$ as defined in (\ref{J-prime}),
and get to a set of evolution equations that are nonsingular
everywhere, even at $\theta_{13}=0$.
The RG evolution equation for $J_{\rm CP}$ and $J'_{\rm CP}$ are obtained as
\beqa
\dot{J}_{\rm CP} &=& A_J + {\cal O}(\theta_{13}) \; , \\
\dot{J'}_{\rm CP} &=& A_J' + {\cal O}(\theta_{13}) \; ,
\eeqa
with
\beqa
A_J &=& C y_\tau^2 s_{12}^2 c_{12}^2 s_{23}^2 c_{23}^2 \frac{m_3}{\dmsq_{31}} 
\Bigl[ m_1 \sin 2\phi_1 - (1+\zeta) m_2 \sin2\phi_2 \Bigr] \; , \\
A_J^\prime &=& C y_\tau^2 s_{12}^2 c_{12}^2 s_{23}^2 c_{23}^2 \frac{m_3}{\dmsq_{31}} 
\Bigl[ m_1 \cos 2\phi_1 - (1+\zeta) m_2 \cos2\phi_2 - \zeta m_3\Bigr] \; .
\label{AJprime}
\eeqa 

We also choose to write the RG evolution for $\theta_{13}^2$ instead 
of $\theta_{13}$, as is traditionally done. This quantity turns out
to have a nonsingular behaviour at $\theta_{13} = 0$.
Moreover, since $\theta_{13} \geq 0$ by convention, the complete 
information about $\theta_{13}$ lies within $\theta_{13}^2$. Also, 
the possible ``sign problem''\footnote{Usually 
the convention used in defining the elements of $U_{\rm PMNS}$ is to take 
the angles $\theta_{ij}$ to lie in the first quadrant. 
${\cal U}_{e3}$ can then take 
both positive or negative values depending on the choice of the CP phase 
$\delta$.  In the formulation of eq. (10) the sign of $A_{13}$ can be such that 
$\theta_{13}$ can assume negative values during the course of evolution
and in such situations one will have to talk about the evolution of 
$|\theta_{13}|$.
Our formulation in terms of 
$\theta_{13}^2$, as shown in eq.~(\ref{th13sq-dot}), 
naturally avoids this problem.} 
of $\theta_{13}$ is avoided.
In terms of the new parameters $J_{\rm CP}$ and $ J_{\rm CP}^\prime$, 
the RG evolution equations for $\theta_{13}^2$ becomes
\beqa
\dot{{\theta_{13}^2} } &=& 
A_{13}^{sq} 
+  {\cal O}(\theta_{13}^2) \; , \label {th13sq-dot} \\
A^{sq}_{13} &=& 8 C y_\tau^2  \frac{ m_3}{\dmsq_{31}}  \Bigl\{  J_{\rm CP}
\left[ m_1 \sin{2 \phi_1} - (1+\zeta) m_2 \sin{2 \phi_2}\right] \Bigr. \nn \\
&& \Bigl. + J_{\rm CP}'
\left[ m_1 \cos{2 \phi_1} - (1+\zeta) m_2 \cos{2 \phi_2} - \zeta m_3\right]  
\Bigr\} \label{A13sq} \; .
\eeqa 
Thus the evolution equations in basis ${\cal P}_J$ are all non-singular 
and continuous at every point. 
In particular, even when $\delta$ shows a discontinuity,
$J_{\rm CP}$ as well as $J_{\rm CP}^\prime$ change in a continuous manner.

\begin{figure}
\parbox{3in}{
\epsfig{file=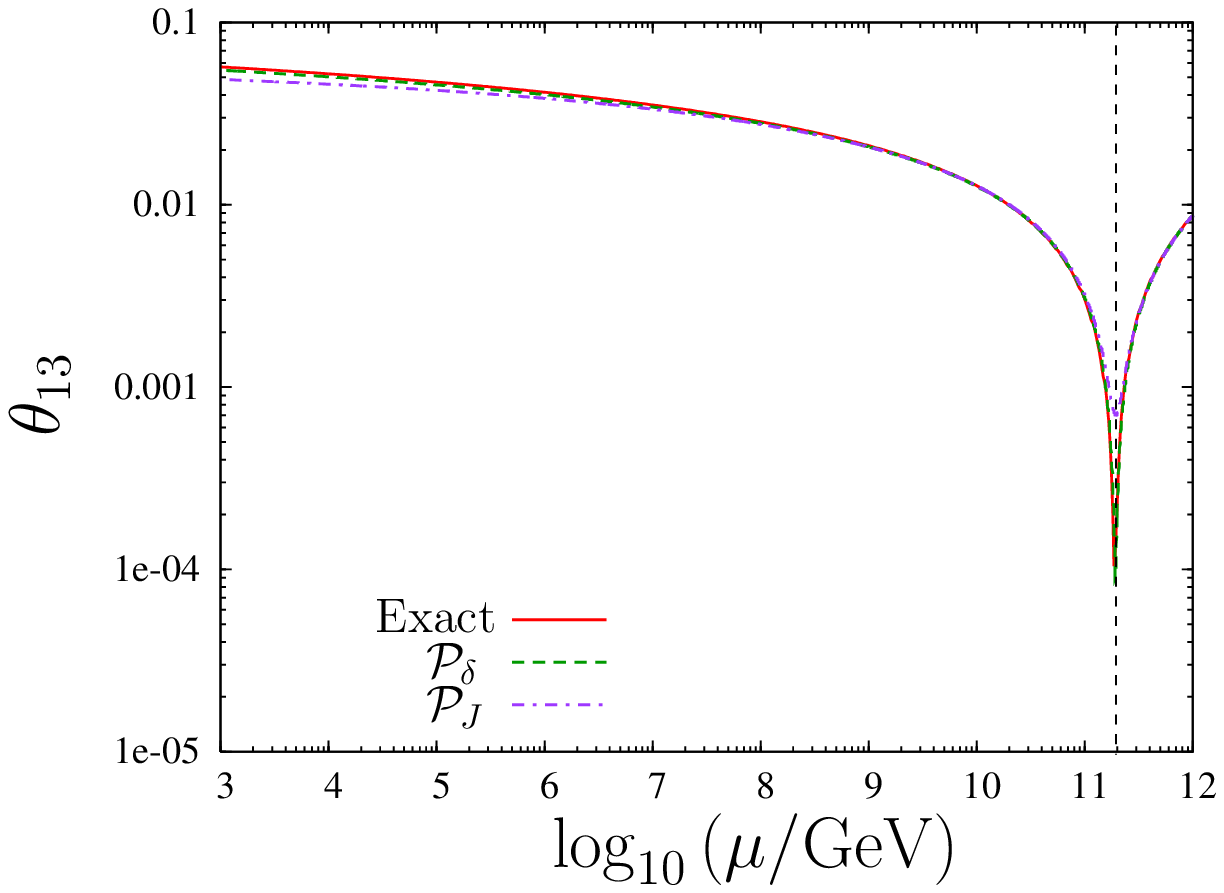,width=3in}
}
\parbox{3in}{
\epsfig{file=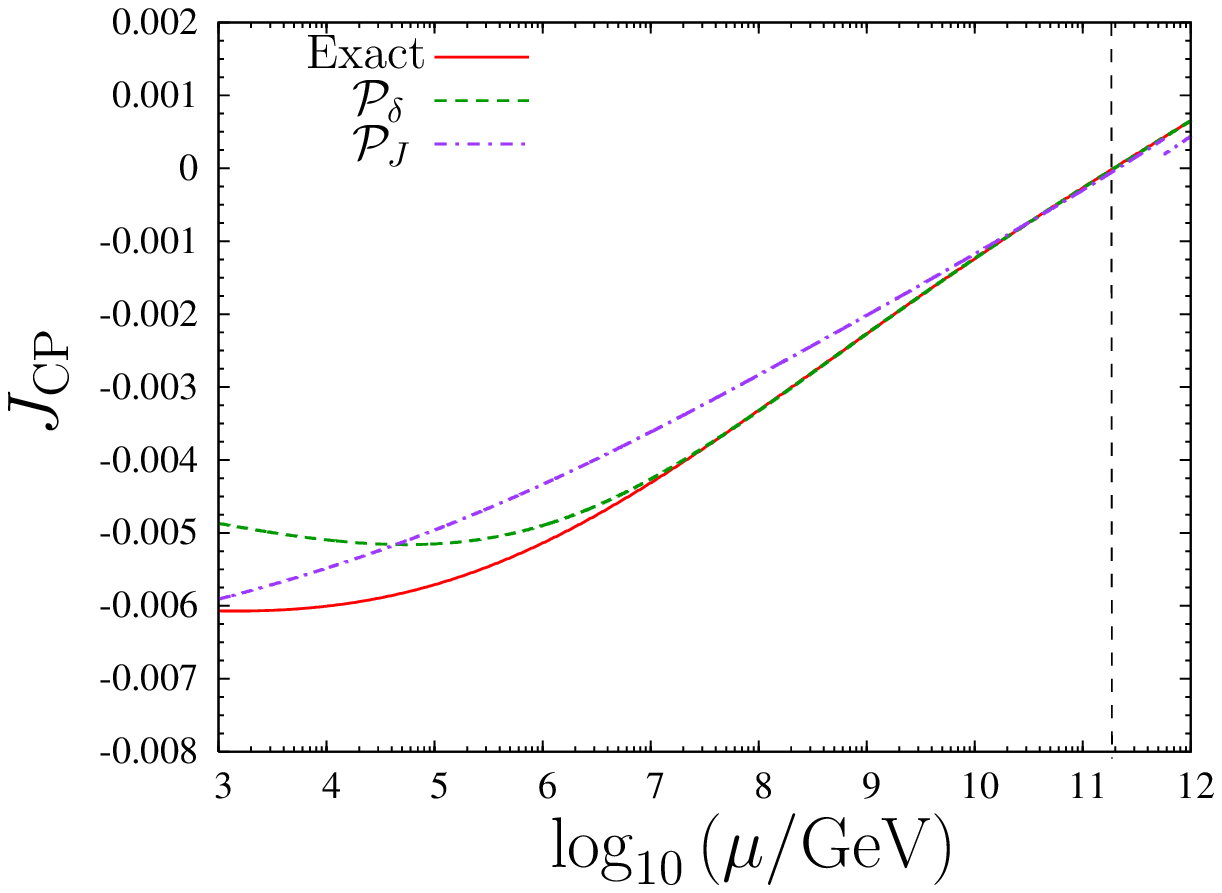,width=3in}
}
\caption{Comparison of the RG evolution of $\theta_{13}$ 
and $J_{\rm CP}$ from the analytic expressions in ${\cal P}_\delta$ 
basis (green, dashed line) and ${\cal P}_J$ 
basis (blue, dotted line) with the exact numeric one (red, solid line). 
The parameters chosen at the high scale $\mu_0 = 10^{12}$ GeV are: 
$\tan{\beta} = 50$, $m_0 = 0.05$ eV$^2$, 
$\dms = 0.00008$ eV$^2$, $\dma = 0.0026$ eV$^2$, 
$\theta_{12} = 34.5^\circ$, $\theta_{23} = 42.5^\circ$ and 
$\theta_{13} = 0.5^\circ$. The phases are taken to be 
$\delta = 40^\circ$, $\phi_1 = 25^\circ$ and $\phi_2 = 105^\circ$. 
\label{th13-jcp-comparison}
}
\end{figure}

In Fig.~\ref{th13-jcp-comparison}, we show the RG evolution of 
$\theta_{13}$ (left panel) and $J_{\rm CP}$ (right panel), as 
obtained from the analytic expressions in ${\cal P}_\delta$ basis 
as well as in the ${\cal P}_J$ basis, along with the 
exact numerical solution, for some chosen values of parameters. 
It shows that the approximate running equations agree with each other
to an accuracy of ${\cal O}(\theta_{13})$.

\section{Bounds on $\theta_{13}$ at low scale} 
\label{bounds}

We now consider all the theories that predict $\theta_{13} = 0$ 
at the high scale and 
try to see the nature of running of the masses and mixing parameters 
with the energy scale. 
For high scale we consider $\mu_0 = 10^{12}$ GeV and 
implement the symmetry $\theta_{13} = 0$ at this scale,
which we also take to be the mass of the lightest heavy
particle responsible for the seesaw mechanism.
We choose this value of $\mu_0$ since it is consistent with the 
current neutrino mass squared differences and seesaw mechanism
with Dirac mass of the heaviest neutrino around $1$--$100$ GeV
\cite{mohapatra-scale}. This scale is also desirable for successful 
leptogenesis  \cite{buchmuller-scale}. 
However, our results are only logarithmically 
sensitive to this choice and hence our conclusions will be robust against 
variations of $\mu_0$.
Also, this would allow us to compare our bounds with those
obtained in \cite{qlctbm} for specific models like tri-bimaximal mixing
at the high scale.
The values of the other parameters at high energy are chosen such that
their low scale values are compatible with experiments.
For the absolute mass scale of neutrinos, we take the cosmological
bound of $m_0 \lesssim 0.5$ eV \cite{hannestad} 
at the laboratory energy.

We consider the scenarios where the effective theory below $\mu_0$
is the SM or the MSSM.
We then estimate the maximum value that $\theta_{13}$ can gain through 
radiative corrections. This can be obtained from
\beqa
\theta_{13} &\equiv& \left\arrowvert \int_{t_0}^{t} A_{13} dt 
+ {\cal O}(\theta_{13}) \right\arrowvert  \\
& \approx& \frac{|C|  \Delta_\tau}{2} \sin{2 \theta_{12}} 
\sin{2 \theta_{23}} \frac{m_3}{|\dmsq_{31}|} \times \nn \\
&& \left\arrowvert m_1 \cos{(2 \phi_1 - \delta)}  - (1+\zeta) m_2 
\cos{(2 \phi_2 - \delta)}  - \zeta m_3 \cos{\delta} \right\arrowvert 
+ {\cal O}(\Delta_\tau \theta_{13},\Delta_\tau^2)
\label{th13-estimate}\; ,
\eeqa
where $t_0 \equiv \ln(\mu_0/{\rm GeV})/(16 \pi^2)$,
$C = -3/2$ for SM and $C = 1$ for MSSM. 
Note that we can use the parameter set ${\cal P}_\delta$ here 
since apart from the starting point, where 
$\delta$ is unphysical and hence is irrelevant completely,
the evolution in terms of this set is also continuous everywhere.
Moreover it is convenient to talk about Dirac and Majorana phases
while putting bounds on quantities.
In eq.~(\ref{th13-estimate}), $\Delta_\tau$ is defined as
\beq
\Delta_\tau^{{\rm SM}} \equiv-\frac{1}{32 \pi^2} 
\left(\frac{g_2 m_\tau}{M_W}\right)^2
\ln{\left( \frac{\mu_0}{\mu} \right)}
\label{Delta_tau-SM}
\eeq
in the SM, where $g_2$ is the SU(2)$_L$ gauge coupling, whereas
$m_\tau$ and $M_W$ are the $\tau$ lepton and W boson 
masses respectively. In the MSSM,
\beq
\Delta_\tau^{{\rm MSSM}} \equiv-\frac{1}{32 \pi^2} 
\left(\frac{g_2 m_\tau}{M_W}\right)^2 (1+\tan^2{\beta}) 
\ln{\left( \frac{\mu_0}{\mu} \right)} \; .
\label{Delta_tau-MSSM}
\eeq
Numerically, one has $\Delta_\tau^{\rm SM} \approx -1.4 \times 10^{-5}$ and
$\Delta_\tau^{\rm MSSM} \approx -1.4 \times 10^{-5} (1 + \tan^2 \beta)$,
where $\tan \beta$ can take values upto $\sim 50$, and
so one can treat these quantities as small parameters.
We explicitly indicate the neglected powers of these parameters
in eq.~(\ref{th13-estimate}).

In order to get the maximum $\theta_{13}$ value possible,
for any value of the lowest neutrino mass $m_0$, all the coefficients of 
the masses $m_i$ in eq.~(\ref{th13-estimate}) should have the same sign
(which we choose to be positive) and the maximum possible magnitude.
This can be achieved with the choice 
\beq
2\phi_1 - \delta_0 = 0 \; , \quad 
|2\phi_2 - \delta_0| = \pi \; ,
\eeq
which gives us
\beqa
\theta_{13}^{\rm max} &\approx& \frac{|C|  \Delta_\tau}{2} \sin{2 \theta_{12}} 
\sin{2 \theta_{23}} \frac{m_3}{|\dmsq_{31}|} 
 \bigl[ m_1+ (1+\zeta) m_2 + |\zeta  m_3 \cos\delta_0|  \bigr]  \\
&\leq & \frac{|C|  \Delta_\tau}{2} \sin{2 \theta_{12}} 
\sin{2 \theta_{23}} \frac{m_3}{|\dmsq_{31}|} 
 \bigl[ m_1+ (1+\zeta) m_2 + |\zeta| m_3  \bigr] 
\label{th13-max}\; .
\eeqa
The right hand side of eq.~(\ref{th13-max}) corresponds to choosing
the phases shown in Table~\ref{tab-mee} for eq.~(\ref{th13-estimate}). 
As seen, these phases depend 
only on whether the neutrino mass ordering is normal or inverted,
and not on the low energy effective theory (SM or MSSM). 
However, the value itself will indeed depend on the effective
theory considered. Note that in this procedure of bounding $\theta_{13}$,
the actual value of $\delta_0$ did not need to be used, a considerable
simplification achieved at the expense of a small overestimation.
\begin{table}[h!]
 \begin{center}
 \begin{tabular}{ccccccc}
 \hline
& \phantom{spc} & $\delta$ & \phantom{spc} & $\phi_1$ & \phantom{spc} & $\phi_2 \phantom{s}$ \\
\hline
  Normal ordering & &  $\pi$  & & $\pi/2$   & & $0$   \\
  Inverted ordering & &  $0$    & & $0$ & & $\pi/2$  \\
  \hline
 \end{tabular}
 \end{center}
\caption{Phase choices in SM and MSSM that give the maximum 
radiative correction 
for $\theta_{13}$.
 \label{tab-mee} }
 \end{table}

To estimate $\theta_{13}^{\rm max}$ that can be generated at the low scale, 
we take the optimal values of the other quantities  
in their current 3$\sigma$ allowed ranges \cite{pdg}.
We are allowed to do this since the corrections to $\theta_{13}$
due to the evolutions of the other quantities will formally be 
${\cal O}(\Delta_\tau^2)$ \cite{rgqlc}.
The quantity that may run quite a bit is $\theta_{12}$, 
however the running is extremely small in the SM
and $\theta_{12}$ always increases in the MSSM, so we use the
maximum allowed value of $\sin 2\theta_{12}$ in eq.~(\ref{th13-max})
for our estimation.
The values of 
$m_1$, $m_2$ and $m_3$ depend on $\dms$, $\dma$, $m_0$ as well 
as the chosen mass ordering. The running of masses and the 
mass squared differences are governed by 
the Yukawa couplings of up-type quarks and 
the U(1)$_Y$ and SU(2)$_L$ gauge couplings. 
For SM, these evolutions depend also on the Higgs boson self coupling, and 
Yukawa couplings of down-type quarks and charged leptons. 
But $\theta_{13}$, as given in eq.~(\ref{th13-max}), will be independent 
of these quantities to the leading order in $\Delta_\tau$ 
and thus considering $\dms$, $\dma$ in the current 3$\sigma$ 
range is expected to give the correct estimate to this order. 
This assumption can be seen to be valid {\it a posteriori} 
from the comparison between analytic and numerical 
results that follow.

\subsection{$\theta_{13}$ at the low scale in the SM}

\begin{figure}[ht!]
\begin{center}
\includegraphics[scale=0.75]{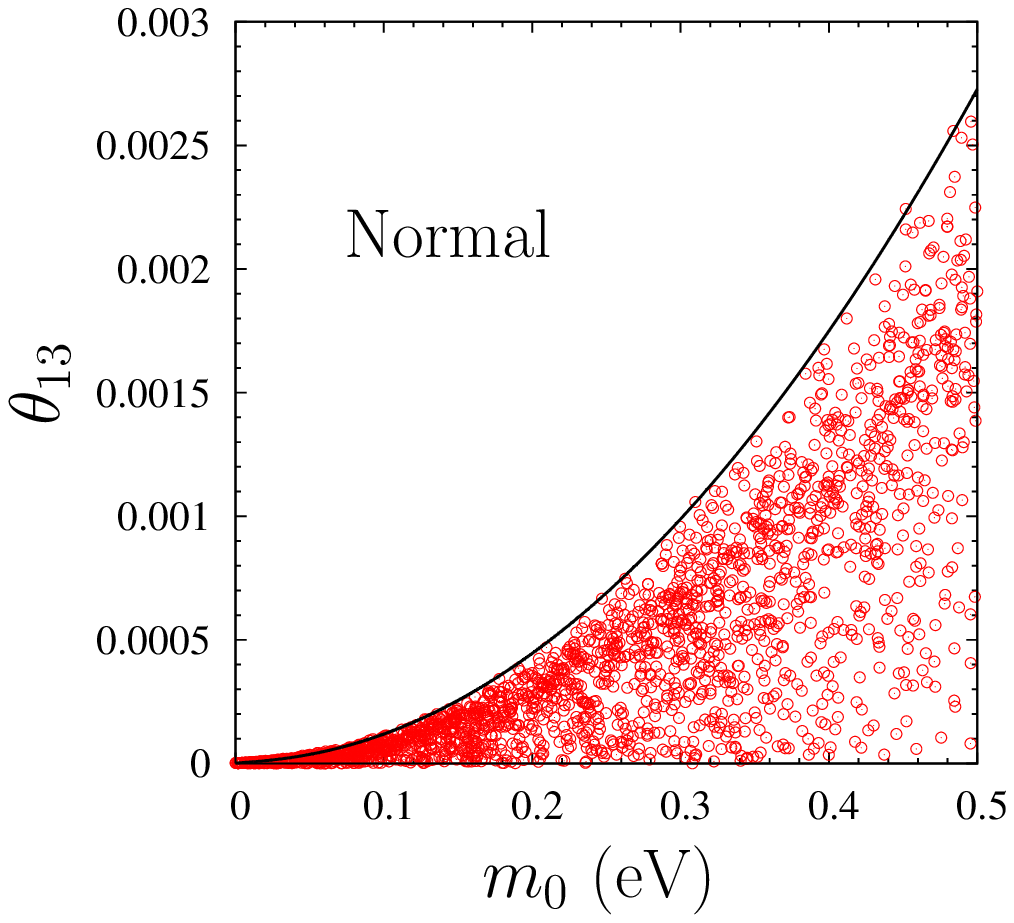}
\includegraphics[scale=0.75]{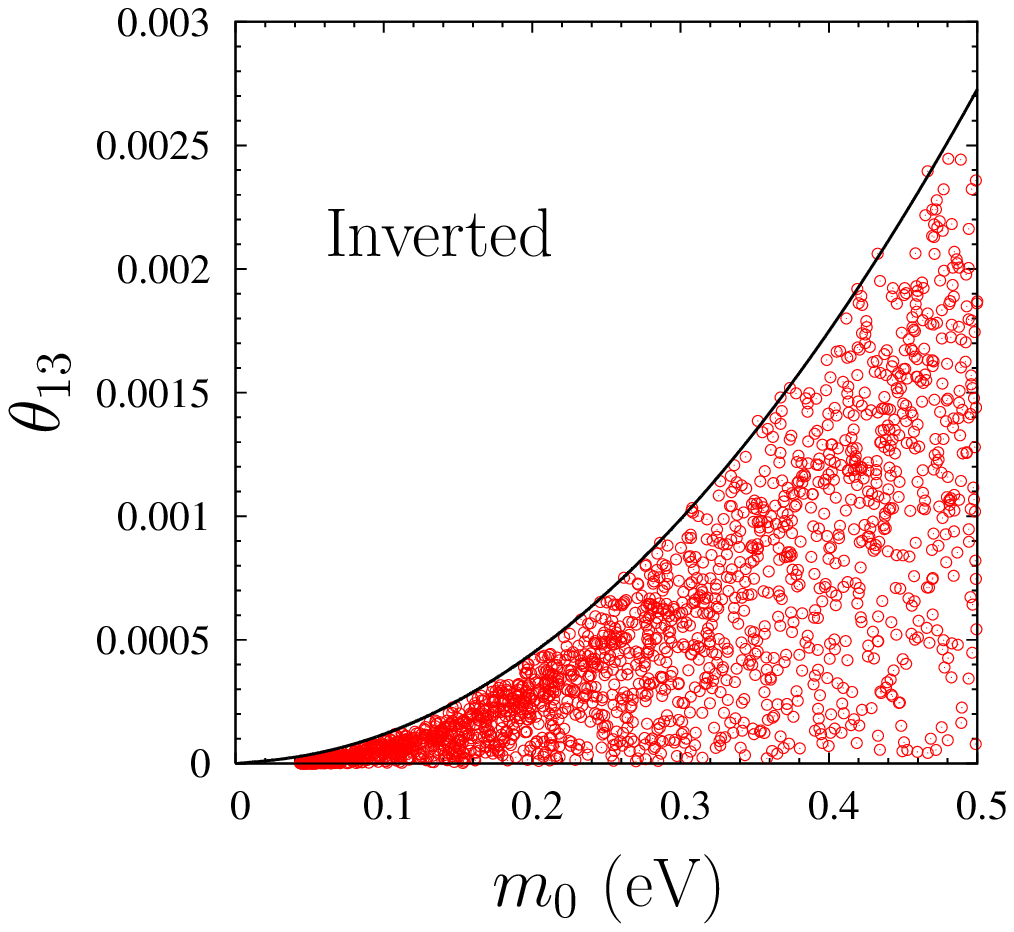}
\caption{Scatter points show the low energy $\theta_{13}$ 
as a function of the lightest neutrino mass $m_{0}$ at the low scale, 
for both normal (left panel) and inverted (right panel) 
mass ordering. Each point represents a 
different high energy theory with $\theta_{13}=0$.
The solid (black) line gives the maximum 
attainable $\theta_{13}$ for a given $m_0$, 
calculated using the analytic bound in eq.~(\ref{th13-max}),
the current 3$\sigma$ limits of the masses and mixings, 
and the  phase values as given in Table~\ref{tab-mee}.
\label{m0-th13-SM}}
\end{center}
\end{figure}
We first consider the case when the effective low energy theory 
below $\mu_0$ is the SM.
Running of the masses and mixing parameters is considered from 
$\mu_0 = 10^{12}$ GeV to the current 
experimental scale ($\sim M_Z$). 
The scatter points in Fig.~\ref{m0-th13-SM} are obtained by keeping 
$\theta_{13} =0$ and varying the other two mixing angles randomly 
in the range 0 to $\pi/2$, whereas the phases are varied between 
0 to $2\pi$. 
The masses at the high scale are varied within $0$--$1.0$ eV, 
so that the lightest neurtino mass $m_0$ at the low 
scale varies between $0$ and $0.5$ eV.
Thus each point represents a different high energy 
theory with $\theta_{13}=0$ at the high scale. 
The upper bound  can be analytically estimated through
eq.~(\ref{th13-max}), which depends on the 
neutrino mass ordering through the phase choices made in 
Table~\ref{tab-mee} and the value of $\Delta_\tau^{SM}$ 
is given in eq.~(\ref{Delta_tau-SM}).

From Fig.~\ref{m0-th13-SM} it is seen that the maximum value
gained radiatively by $\theta_{13}$ is rather small,
being $\lesssim 3 \times 10^{-3}$ 
in the range $0 \leq m_{0} \leq 0.5$ eV
for both the mass orderings. 
Hence if future experiments 
measure $\theta_{13}$ greater than this limit,
all the theories with $\theta_{13} = 0$ at the high scale and
SM as the low energy effective theory will be ruled out completely.
If the upper limit for $m_0$ is brought down by KATRIN \cite{katrin}
to $m_0 \lesssim 0.2$ eV, even lower $\theta_{13}$ values will be
excluded for this class of theories. Note that for $m_0$ 
of this order, the effective electron neutrino mass measured 
by KATRIN will essentially be the same as $m_0$.

\subsection{$\theta_{13}$ at the low scale from MSSM}

%
\begin{figure}[ht!]
\begin{center}
\includegraphics[scale=0.8]{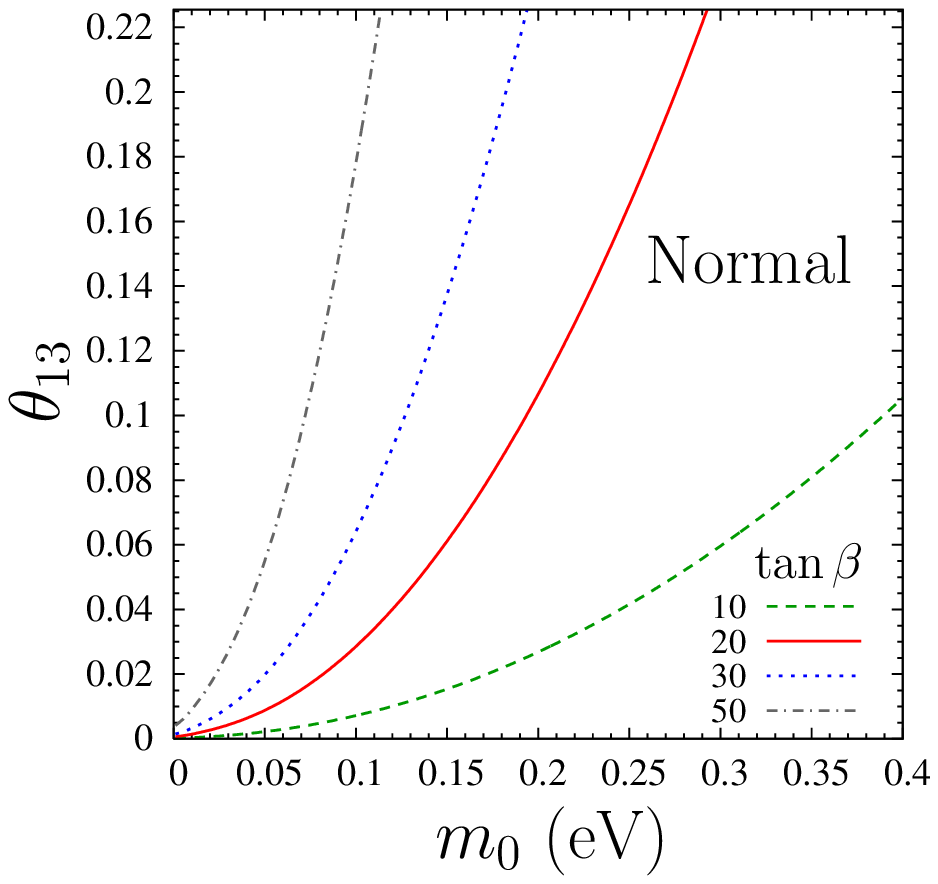}
\includegraphics[scale=0.8]{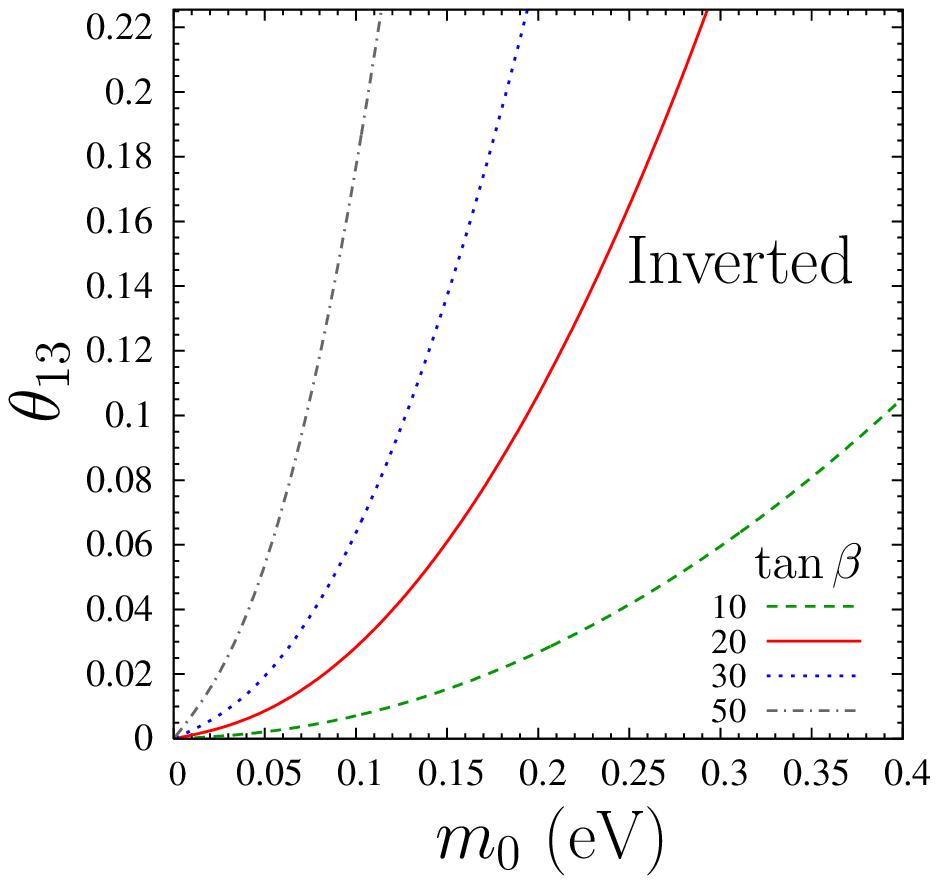}
\caption{Maximum $\theta_{13}$ obtained at the low scale as a 
function of the lightest neutrino mass $m_0$ at the low scale 
for $\tan{\beta}$ = $10$, $20$, $30$ and $50$
in the normal (left panel) and 
inverted (right panel) mass ordering. The plots 
show that simultaneous measurement of $\theta_{13}$ and 
$m_0$ will help in ruling out of a class of high energy 
theories with $\theta_{13} = 0$. However there is a 
strong dependence on the upper limit of $\tan{\beta}$.
\label{m0-th13}}
\end{center}
\end{figure}

When MSSM is the low energy effective theory, the evolution of the
neutrino parameters is proportional to $(1 + \tan^2 \beta)$, 
as is seen from eq.~(\ref{Delta_tau-MSSM}),
where $\tan\beta$ may take values up to $\sim 50$.
Thus, considerably larger running of $\theta_{13}$ may be expected
at large $\tan\beta$.
The variation of $\theta_{13}$ as a 
function of $m_0$ is shown in Fig.~\ref{m0-th13}.
From the figure it can be concluded that
with the current limit of $m_0$, the radiative correction to 
$\theta_{13} = 0$ at the high scale can be large enough to reach 
the present upper bound of $\theta_{13}$ at laboratory energy. 
However, for a given $m_0 \lesssim$ 0.1 eV, the maximum 
$\theta_{13}$ these theories can generate is significantly lower 
for the whole $\tan{\beta}$ range. For example, if $m_0$ happens 
to be $0.08$ eV, the maximum $\theta_{13}$ for $\tan\beta=50$ is 
$\theta_{13} \sim 0.12$, i.e. $\sin^2{2 \theta_{13}} \sim 0.056$. 
Such a $\theta_{13}$ 
regime will be probed by the next generation 
neutrino oscillation experiments like Double CHOOZ \cite{DCHOOZ}, 
Daya Bay \cite{daya-bay}, T2K \cite{t2k}. 
Since the tritium beta decay experiment KATRIN \cite{katrin} 
plans to probe $m_0 \sim 0.2$ eV only, it may not be enough 
to rule out theories with larger $\tan\beta$.

However, the neutrinoless double beta decay 
($0\nu\beta\beta$) experiments will measure the 
effective Majorana mass of the electron neutrino
\beqa
m_{ee} = \left| c_{12}^2 c_{13}^2 m_1 e^{2 i (\phi_1 - \delta)} + 
s_{12}^2 c_{13}^2 m_2 e^{2 i (\phi_2 - \delta)} + 
s_{13}^2 m_3 \right| \; ,
\label{mee}
\eeqa
The value of $m_{ee}$ will allow us 
to estimate the $m_0$ range, albeit with 
a large uncertainty owing to the complete lack of 
knowledge of the phases $\delta$, $\phi_1$ and $\phi_2$ 
currently. The present upper bound on the average neutrino mass 
is $m_{ee} < 1.1$ eV \cite{heidelberg-moscow}, whereas
the proposed next generation experiments like
COBRA\cite{cobra}, CUORE \cite{cuore}, EXO\cite{exo}, 
GERDA \cite{gerda}, Super-NEMO\cite{s-nemo}, MOON \cite{moon}
plan to probe $m_{ee}$ in the range as low as
$0.01 {\; \rm eV} \leq m_{ee} \leq 0.1 {\; \rm eV}$. 
Therefore, combined measurement of $\theta_{13}$ and 
$m_{ee}$ may enable us to put some bound on the 
theories with large $\tan \beta$.

\begin{figure}[t!]
\begin{center}
\includegraphics[scale=0.8]{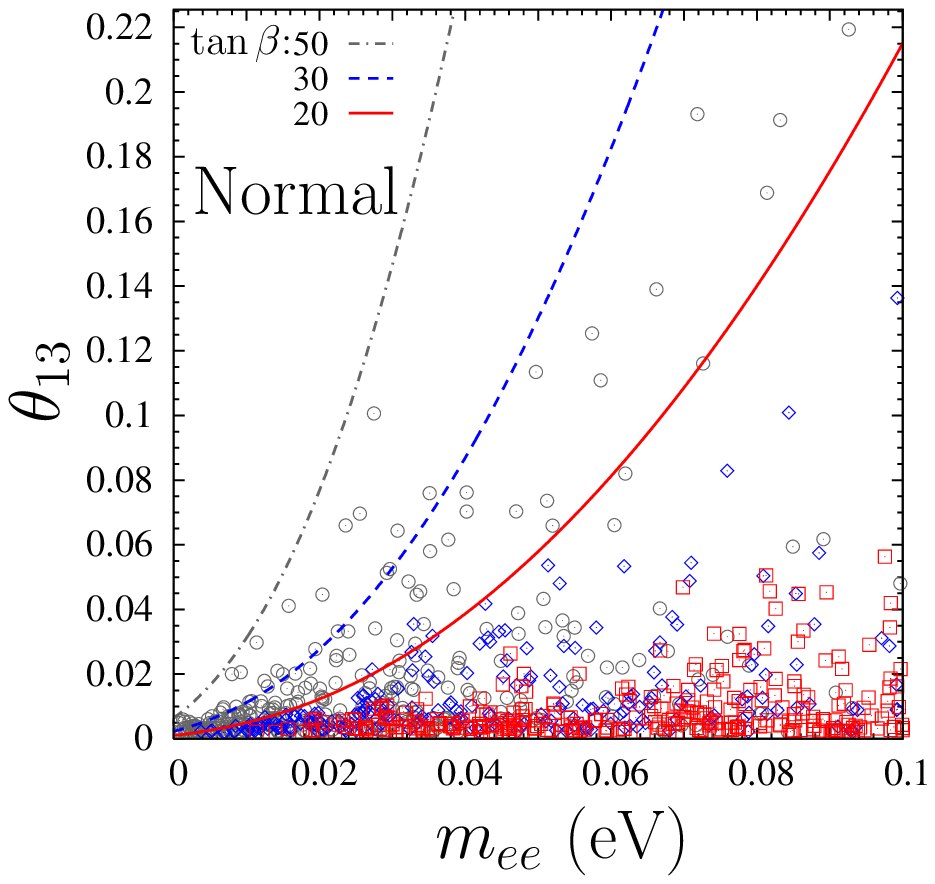}
\includegraphics[scale=0.8]{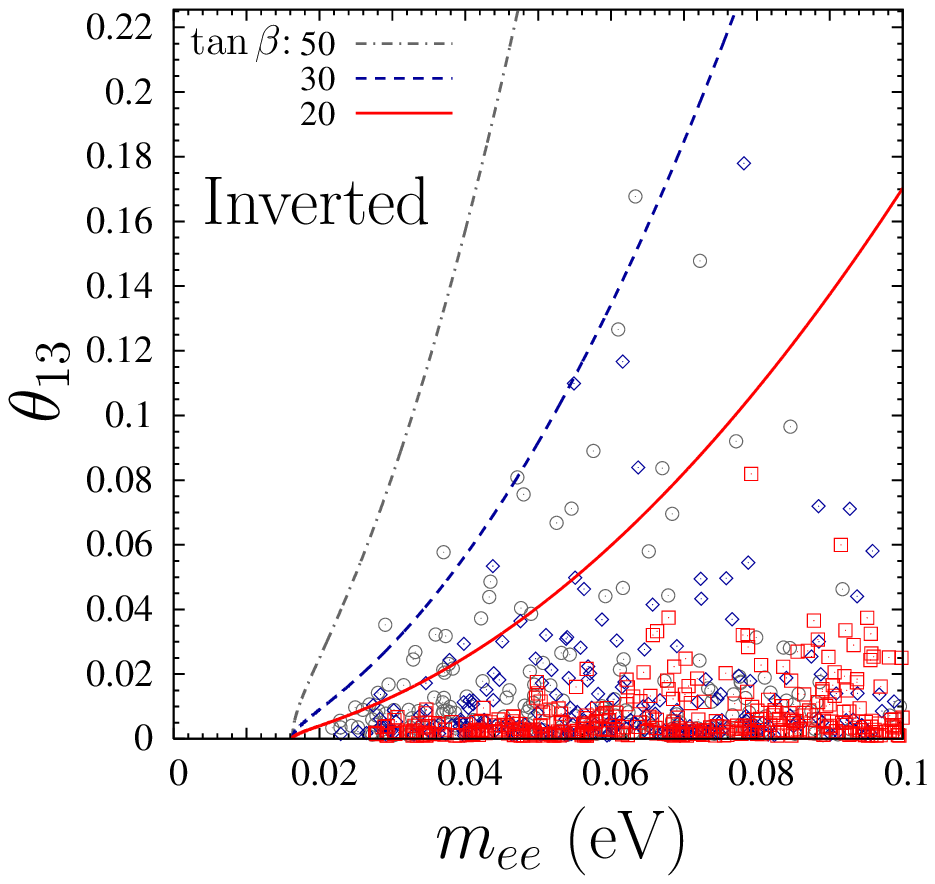}
\caption{Scatter points show the value of $\theta_{13}$ generated
at the low scale as a function of $m_{ee}$,
for normal (left panel) and inverted (right panel) ordering.
Each point represents a different high energy theory with 
$\theta_{13}=0$.
Different symbols (colors) correspond to different ranges of 
$\tan{\beta},$ {\it viz.} 
squares (red) for $1.0 \leq \tan{\beta} \leq 20.0$, 
diamonds (blue) for $20.0 \leq \tan{\beta} \leq 30.0$ 
and circles (gray) for $30.0 \leq \tan{\beta} \leq 50.0$. 
The lines show analytic estimates of $\theta_{13}^{\rm max}$:
solid (red) line for $\tan{\beta} = 20.0$, dashed (blue) 
line for $\tan{\beta} = 30.0$ and dot-dashed (gray) line for 
$\tan{\beta} = 50.0$. 
\label{mee-th13-MSSM}}
\end{center}
\end{figure}

The expression for $m_{ee}$ in (\ref{mee}) can be expanded
in terms of the 
parameter $\delta_\odot \equiv \dms/m_0^2$, which 
is small in the range $m_{ee} > 0.01$ eV, 
and the small parameter $\theta_{13}$, to get
\beqa
m_{ee} = m_0 \cos{2\theta_{12}} \left( 1 - \frac{\delta_\odot}{2} 
\frac{s_{12}^2}{\cos{2\theta_{12}}} 
- \theta_{13}^2  \right)-  \theta_{13}^2  \sqrt{m_0^2+\dma} 
+ {\cal O}(\delta_\odot^2,\delta_\odot \theta_{13}^2,\theta_{13}^3) \; 
\label{mee-m0-NH}
\eeqa
for normal mass ordering,
where the phases are chosen as given in Table~\ref{tab-mee}.
For inverted mass ordering,
\beq
m_{ee} = \cos{2 \theta_{12}} (1-\epsilon) \sqrt{m_0^2+ |\dma|} \; 
+ {\cal O}(\delta_\odot^2,\delta_\odot \theta_{13}^2,\theta_{13}^3) \; ,
\label{mee-m0-IH}
\eeq
where
\beq
\epsilon = \frac{\delta_\odot}{2} \frac{c_{12}^2}
{ \cos{2 \theta_{12}} (1 + \Delta)} 
+ \theta_{13}^2 \left( 1 - \frac{1}{\cos{2 \theta_{12}} 
\sqrt{1+\Delta}} \right) \; .
\eeq
The quantity $\Delta \equiv |\dma|/m_0^2$ is bounded from below, 
while for inverted mass ordering $\delta_\odot$ is a small parameter
($\sim {\cal O}(10^{-1})$) in the range $m_{ee} > 0.02$ eV, 
so that $\epsilon$ is small in this range. 
The analytic expressions in eqs.~(\ref{mee-m0-NH}) and 
(\ref{mee-m0-IH}) are valid for  
$m_{ee} > 0.01$ eV and 
$m_{ee} > 0.02$ eV respectively. In this domain of validity, 
we invert the relations (\ref{mee-m0-NH}) and (\ref{mee-m0-IH})
to obtain $m_0$ in terms of $m_{ee}$, and then use 
eq.~(\ref{th13-max}) for an analytic estimation 
of $\theta_{13}^{\rm max}$.
For the $m_{ee}$ values outside the range of validity, 
one has to estimate 
numerically the minimum allowed $m_{ee}$ for a given $m_0$ 
and then use eq.~(\ref{th13-max}) to 
determine $\theta_{13}^{\rm max}$.
These estimations are shown in Fig.~\ref{mee-th13-MSSM}
for various $\tan\beta$ values.
The scattered points are the low scale predictions calculated
numerically, which show the correlated constraints in the 
parameter space of $\theta_{13}$ and $m_{ee}$.
It may be noted that the analytic bounds on $\theta_{13}$ 
obtained here as a function of $m_{ee}$ are generous overestimations, 
mainly due to the error in the estimation of $m_0$ for a given $m_{ee}$.

Note that bounds on $\theta_{13}$ at the low scale 
generated by RG evolution have been studied earlier in the context of
specific neutrino mixing scenarios at the high scale, like the 
quark-lepton complementarity or tri-bimaximal mixing \cite{qlctbm},
or correlated generation of $\dmsq_{21}$ and $\theta_{13}$ 
\cite{Joshipura:2002kj,Joshipura:2002gr}.
The bounds obtained in this section, which are applicable not only 
for all the models with $\theta_{13}=0$ at the high scale,
but to all the models with $\theta_{13}=0$ anytime during their RG
evolution, subsumes the earlier analyses with specific models.

\section{Summary}
\label{summary}

If the neutrino mixing angle $\theta_{13}$ is extremely small, 
it could point towards some flavor symmetry in the lepton sector.
There is indeed a large class of theories of neutrino mass that 
predict extremely small or even vanishing $\theta_{13}$.
However, such predictions are normally valid at the high scale
where the masses of the heavy particle responsible for neutrino
mass generation lie.
Below this scale, radiative corrections give rise to RG evolution
of the neutrino mixing parameters, which in principle can wipe
out signatures of such symmetries.
In this paper, we explore the RG evolution of all such theories 
collectively.

The RG evolution with the traditional parameter set  
${\cal P}_\delta = \{m_i, \theta_{ij}, \phi_i, \delta\}$ 
involves an apparent singularity in the evolution of the 
Dirac phase $\delta$ when $\theta_{13} = 0$.
This singularity is unphysical, since all the elements of
the neutrino mixing matrix $U_{\rm PMNS}$ are continuous at $\theta_{13}=0$,
and in fact the value of $\delta$ there should be immaterial.
A practical solution to this situation has already been proposed,
which involves prescribing a specific value of $\cot\delta$
when one starts the RG evolution of a model with $\theta_{13}=0$.
However, the rationale behind this prescription has not been
studied before in the context of the nature of this apparent 
singularity.
This issue also becomes relevant for the class of models under
consideration here, since if $\theta_{13}$ is very close to zero at
the high scale, it may vanish completely during its RG evolution,
and getting the required value of $\delta$ exactly at that point
seems like fine tuning.

We explore the apparent singularity in $\delta$ by analyzing the
evolution of the complex quantity ${\cal U}_{e3}$, which stays continuous
throughout the RG evolution.
We find that a fine tuning is indeed required, but that is to 
ensure that $\theta_{13}$ exactly vanishes.
In general, if the CP violating Dirac and Majorana phases take 
nontrivial values, one does not pass through $\theta_{13}=0$
even when one starts with $\theta_{13}$ very close to zero.
One needs rather finely tuned values for the starting values
of the neutrino mixing parameters, unless one introduces a
symmetry like CP conservation, which makes the Dirac and
Majorana phases vanish everywhere.
Since the latter assumption is used commonly in literature,
one tends to miss the fact that getting $\theta_{13}=0$
during RG evolution is possible only in a small region of the
parameter space.

However, if the parameters happen to be tuned such that 
$\theta_{13}$ vanishes exactly, we show that the limiting value
of $\delta$ as $\theta_{13} \to 0$ is indeed the one given
by the prescription mentioned above. We thus put the 
prescription on a solid footing by deriving it from first
principles. 
We also propose an alternate parametrization using the parameter set
${\cal P}_J = \{m_i, \theta_{12}, \theta_{23}, \theta_{13}^2, \phi_i, 
J_{\rm CP}, J'_{\rm CP} \}$, where all the parameters are 
well-defined everywhere and
any seemingly nonsingular behavior is avoided.

For models with exactly vanishing $\theta_{13}$ at the high scale,
we study the generation of nonzero $\theta_{13}$ 
through radiative corrections.
We consider two scenarios, one when the low energy effective theory
is the SM, and the other where it is the MSSM.
The radiatively generated $\theta_{13}$ values are correlated
with the absolute neutrino mass scale $m_0$.
This scale will be probed by the future experiments 
on tritium beta decay, and indirectly by the neutrinoless
double beta decay experiments.
If the value of $m_0$ is indeed restricted to the value $\sim 0.2$ eV 
which KATRIN will probe,
the maximum value of $\theta_{13}$ generated can only
be $\lesssim 3 \times 10^{-3}$  in the SM scenario.
With the MSSM, the running can be much higher for large
$\tan\beta$, such that
the current bound of $\theta_{13} < 0.22$ may be reached.
In this scenario, we correlate the bound on $\theta_{13}$
with the effective neutrino Majorana mass $m_{ee}$ 
to be measured in the next generation neutrinoless double 
beta decay experiments.
The whole class of models considered in this paper can then be
ruled out from future measurements of $\theta_{13}, m_{ee}$ and
$\tan\beta$.

\section*{Acknowledgements}

We thank the organisers of WHEPP-X and Neutrino2008, the conferences 
during which this work started and developed. 
We also thank Probir Roy for useful discussions.
S.R. would like to thank Harish-Chandra Research Institute 
and S.G. would like to thank Tata Institute of 
Fundamental Research for kind hospitality. 
The work of A.D. and S.R. was partially supported
by the Max Planck -- India Partnergroup program between
Max Planck Institute for Physics and Tata Institute of
Fundamental Research. S.G. acknowledges support from the Neutrino
Project under the XI$^{\mbox{th}}$ 
plan of Harish-Chandra Research Institute.


\end{document}